\documentclass[%
 aps,
prl,
superscriptaddress,
 amsmath,amssymb,
reprint,%
]{revtex4-2}

\usepackage{graphicx}
\usepackage{dcolumn}
\usepackage{bm}
\usepackage[utf8]{inputenc}
\usepackage[T1]{fontenc}
\usepackage{mathptmx}
\usepackage{xcolor}

\begin{document}

\preprint{V3}

\title{Wurtzite MnSe as a barrier for CdSe quantum wells with built-in electric field}

\author{M.~J.~Grzybowski}
 \email{Michal.Grzybowski@fuw.edu.pl}
  \affiliation{University of Warsaw, Faculty of Physics, Pasteura 5, 02-093 Warsaw, Poland}
\author{W.~Pacuski}
 \affiliation{University of Warsaw, Faculty of Physics, Pasteura 5, 02-093 Warsaw, Poland}
%
\author{J.~Suffczyński}
 \affiliation{University of Warsaw, Faculty of Physics, Pasteura 5, 02-093 Warsaw, Poland}



\date{\today}

\begin{abstract}
Altermagnetic materials have attracted a lot of attention recently due to the numerous effects, which have an application potential and occur due to the spin-split band structure coexisting with the compensated magnetic order. Incorporation of such intriguing compounds into low-dimensional structures represents an important avenue towards exploiting and enhancing their functionalities. Prominent examples of this group are semiconductors well suited to the band-gap engineering strategies. Here, we present for the first time visible-light-emitting CdSe quantum wells, in which wurtzite MnSe as an alermagnetic candidate plays the role of a barrier. Photoluminescence experiments with temporal resolution demonstrate that in such quantum wells, a built-in electric field is present and strongly influences the energies of the emitted photons, the dynamics of recombination, and excitation power dependence. Numerical simulations allow us to estimate that the magnitude of the electric field is $14\,$MV/m. We anticipate that such quantum wells offer potential to probe the barrier properties and that wurtzite MnSe is an interesting platform to study the interplay of the altermagnetism and built-in electric field.\\
%
\end{abstract}

\maketitle

The combination of the internal electric field and magnetic order in a semiconductor provides a wide variety of interesting electronic, magnetic, and optical properties. Among magnetically doped III-V wurtzite-structure semiconductors, nitrides are well known to exhibit effects related to the presence of an internal electric field \cite{DellaSala1999, Cingolani2000, Reale2003, Sztenkiel2016, Suffczynski2011, Pacuski2008}. Respective II-VI systems involve wurtzite CdSe \cite{Langbein1994, Morello2008, Halsall1992, Ghosh2015} or ZnO \cite{Morhain2005, Pacuski2011}. The internal electric field manifests itself especially strongly in low-dimensional structures in the form of quantum confined stark effect (QCSE) \cite{Miller1984}. Optical studies of such systems can provide important information about the built-in potential gradient \cite{Miller1985}. 

An intensive recent interest in the role of symmetries in magnetic compounds \cite{Yuan2020} led to a novel classification of collinear magnetic materials, in which compounds with compensated magnetic order and spin-splitting of the bands in zero magnetic field \cite{Krempasky2024}, called altermagnets play a very important role \cite{Smejkal2022a, Smejkal2022b}. Numerous interesting effects considered as being relevant only for ferromagnets are now predicted to occur due to the altermagnetism without a net magnetization. They include efficient spin current generation \cite{Gonzalez-Hernandez2021}, and giant or tunneling magnetoresistance \cite{Smejkal2022c}. Diverse materials have been predicted to possess altermagnetic properties recently \cite{Bai2024}. Wurtzite MnSe is one of them. Since its epitaxial growth can be conveniently performed on buffer layers deposited on common GaAs (111) substrates, \cite{Grzybowski2024} it is compatible with epitaxial low-dimensional structures such as II-VI quantum wells (QWs). The symmetry of the wurtzite MnSe is close to that of intensively explored NiAs-type MnTe \cite{GonzalezBetancourt2023, Hariki2024, Osumi2024, Kluczyk2024, Amin2024}. Contrary to such NiAs-type MnTe, wurtzite MnSe is an example of a noncentrosymmetric altermagnet \cite{Autieri2025}, where significant electrical polarization can be expected \cite{Bezzerga2025}. Low-dimensional semiconductor structures such as QWs offer enhanced optical transitions and additional degrees of freedom, such as resonances controllable by QW thickness and barrier band-gaps. As such, they provide an important and highly desirable environment to study magneto-optical effects leading towards read-out and control of the altermagnetic states. In particular, the exchange interaction between the excitons trapped in the QW potential and magnetic moments localized on Mn atoms seems to create appealing opportunities for novel magnetic probing techniques.

\begin{figure}
\includegraphics[width=1\linewidth]{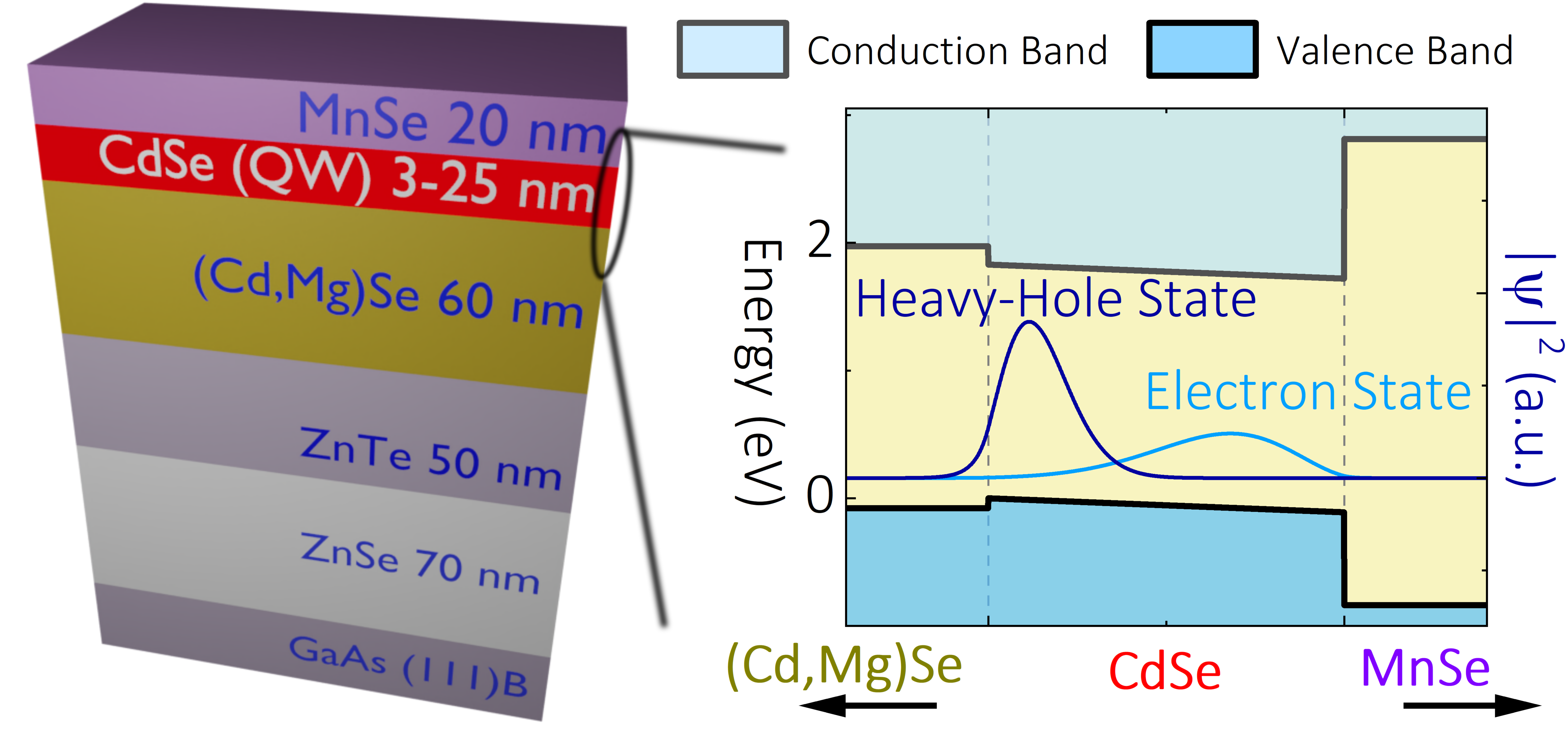}
\caption{\label{fig:schemat}The scheme of the epitaxial, wurtzite-type structures with a CdSe quantum well in MnSe and (Cd,Mg)Se barriers, along with the schematics of their band alignment and the calculated spatial distribution of QW-confined the electron and heavy-hole density in ground states. }
\end{figure}

Wurtzite-structure MnSe has been obtained only recently \cite{Grzybowski2024} and indicated as an altermagnetic candidate. In this report, we demonstrate that epitaxial, wurtzite-structure MnSe can be implemented in a low-dimensional quantum structure as a barrier for asymmetric wurtzite CdSe QWs emitting in the visible spectral range. Analysis of the time-integrated and time-resolved emission of QW-confined exciton as a function of the QW width and excitation power dependence allows us to conclude about the presence and the important role of the built-in electric field in the system. Because MnSe shares the same crystal symmetry as the wurtzite CdSe quantum well, the electric field is also present in the wurtzite MnSe barrier.

All of the samples investigated in this study are epitaxially grown II-VI semiconductor heterostructures on GaAs (111) B substrates with asymmetric CdSe quantum wells. The only varied structural parameter for a set of samples considered is the QW thickness, while all the others remain unchanged. The bottom barrier is (Cd,Mg)Se with around $12 \% $ of Mg, whereas the top barrier is the wurtzite-structure MnSe. ZnSe and CdSe buffer layers are used to mitigate the lattice mismatch effects and to stabilize the wurtzite phase of both CdSe and MnSe as schematically presented in Fig.~\ref{fig:schemat}. The growth temperature is set at $250^{\circ}$ C. Significantly higher temperatures of growth prevent CdSe growth, whereas much lower temperatures cause defects in MnSe, as can be inferred from reflection high-energy electron diffraction (RHEED) patterns. Details of growth and material characterization can be found elsewhere \cite{Grzybowski2024}.

\begin{figure}
\includegraphics[width=1\linewidth]{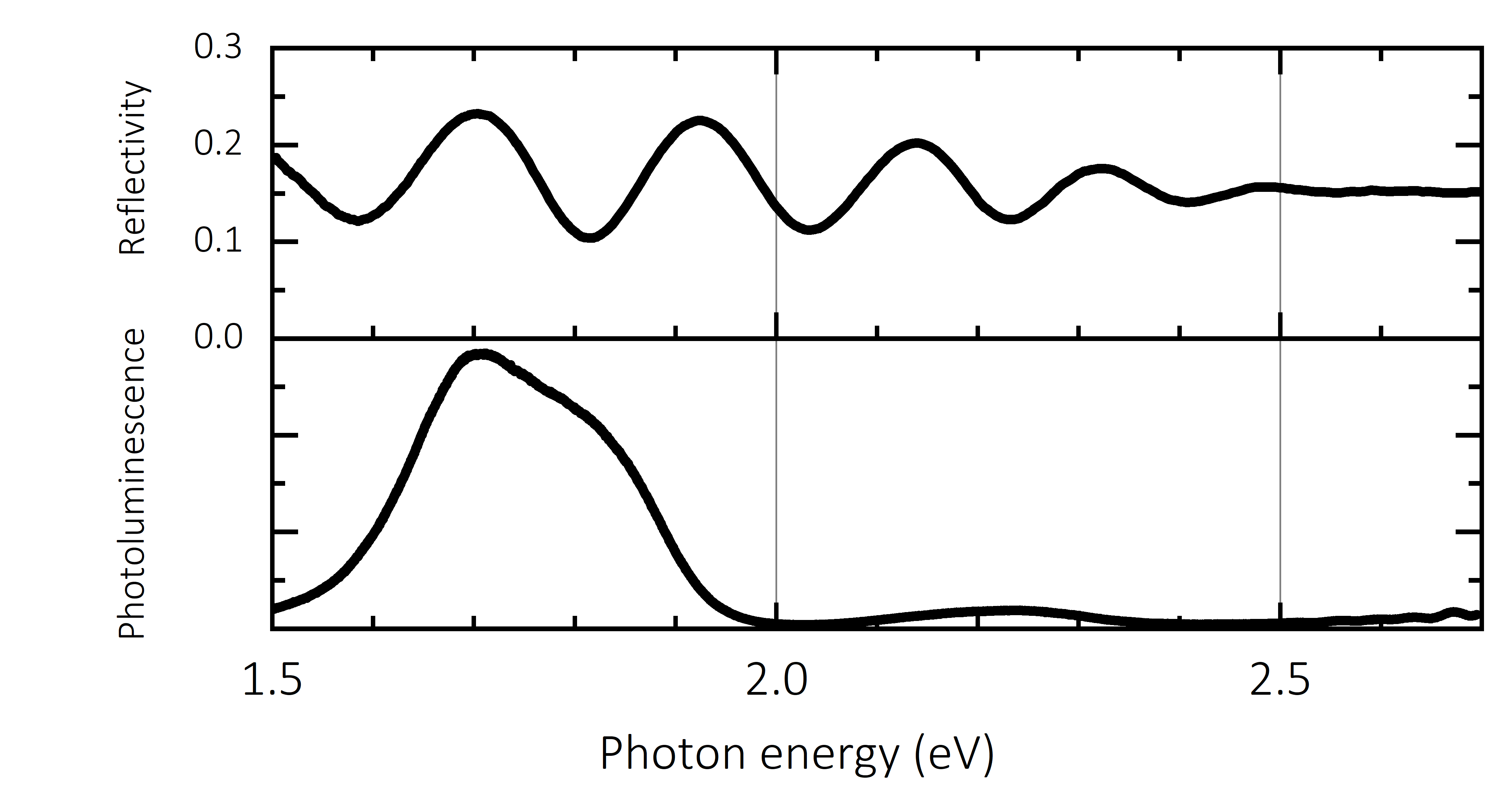}
\caption{\label{fig:bandgap}Reflectivity and photoluminescence spectra collected at room temperature for $1\,\mu$m thick wurtzite MnSe layer grown on CdSe, ZnTe and ZnSe buffers on GaAs (111)B substrate. Photoluminescence is excited with $\lambda_\text{exc}=405\,$nm.} 
\end{figure}

Although thick epitaxial MnSe layers can have a significant admixture of the undesired rock-salt phase as well as possess the neighborhood of the buffer layers, which complicates the analysis, their optical studies illustrate very clearly the difficulty in band gap determination (Fig.~\ref{fig:bandgap}). Reflectivity and photoluminescence spectra for the same sample do not share a common feature, which can be attributed to the interband transitions of the MnSe. The reflectivity spectrum exhibits an interference pattern, typical for a layered material, that dims above $2.5\,$eV, which is related to the absorption and suggests that the band gap is at least $2.5\,$eV. The reflectivity value allows us to estimate that the refractive index of wurtzite MnSe is $n=2.26$ in the considered energy range. Meanwhile, the photoluminescence signal is dominated by a broad maximum centered around $1.7\,$eV, which is most likely related to the intraionic Mn transitions. The band gap of the wurtzite MnSe was theoretically calculated to be around $2\,$eV \cite{Grzybowski2024}. A likely underestimation of this value resulting from ab-initio methods together with much higher values determined in experiments with chemically synthesized nanoparticles \cite{Sines2010} as well as the reflectivity spectra indicating increased absorption in the higher energies of the visible range allows us to expect that wurtzite MnSe can be a good barrier material for CdSe QW, which we further verify in the following.


\begin{figure}
\includegraphics[width=1\linewidth]{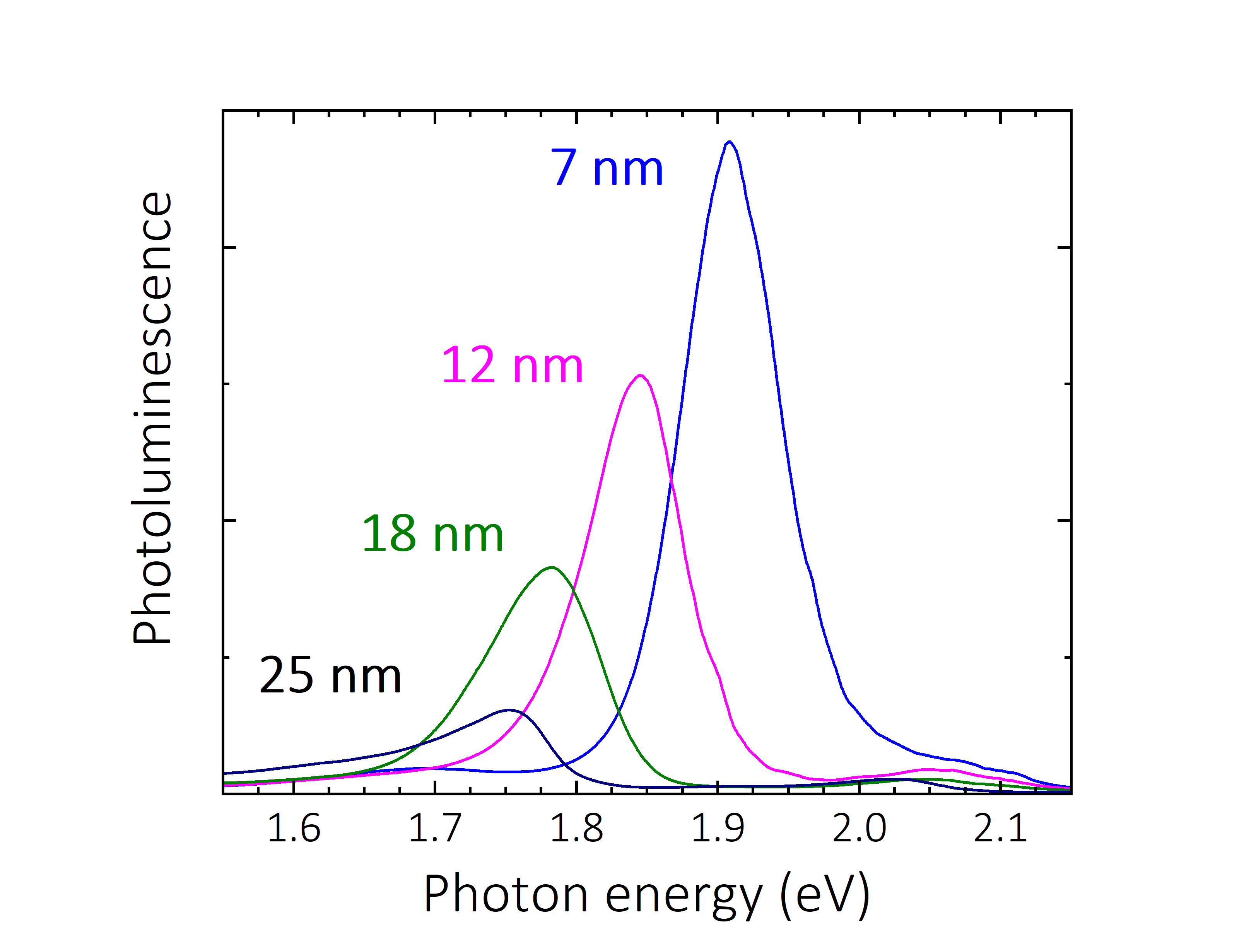}
\caption{\label{fig:widma} Photoluminescence spectra of several samples containing wurtzite CdSe quantum wells differing in the thickness, while the barrier materials - (Cd,Mg)Se and MnSe, as well as other parameters of the structure, remain unchanged. The spectra ar  collected with the excitation power of $2.5\,$mW, wavelength of $\lambda=405\,$nm at $T=10\,$K. }
\end{figure}

Photoluminescence of samples with QWs varying only in thickness (as explained in previous paragraphs and depicted in Fig.~\ref{fig:schemat}) is measured in a He-flow cryostat. Resulting spectra for cryogenic temperatures ($T=10\,$K) with the excitation wavelength $\lambda_\text{exc}=405\,$nm and power of $P=2.1\,$mW are presented in Fig.~\ref{fig:widma}. QW well thicknesses are indicated by the colorful captions. Broad maxima between $1.7\,$eV and $2.0\,$eV represent photoluminescence from the CdSe QW. Maximum intensity of the PL signal is observed for $7\,$nm QW. Remarkably, the emission from the QW is also seen at room temperature, where it has an overall much lower intensity. 

\begin{figure}
\includegraphics[width=1\linewidth]{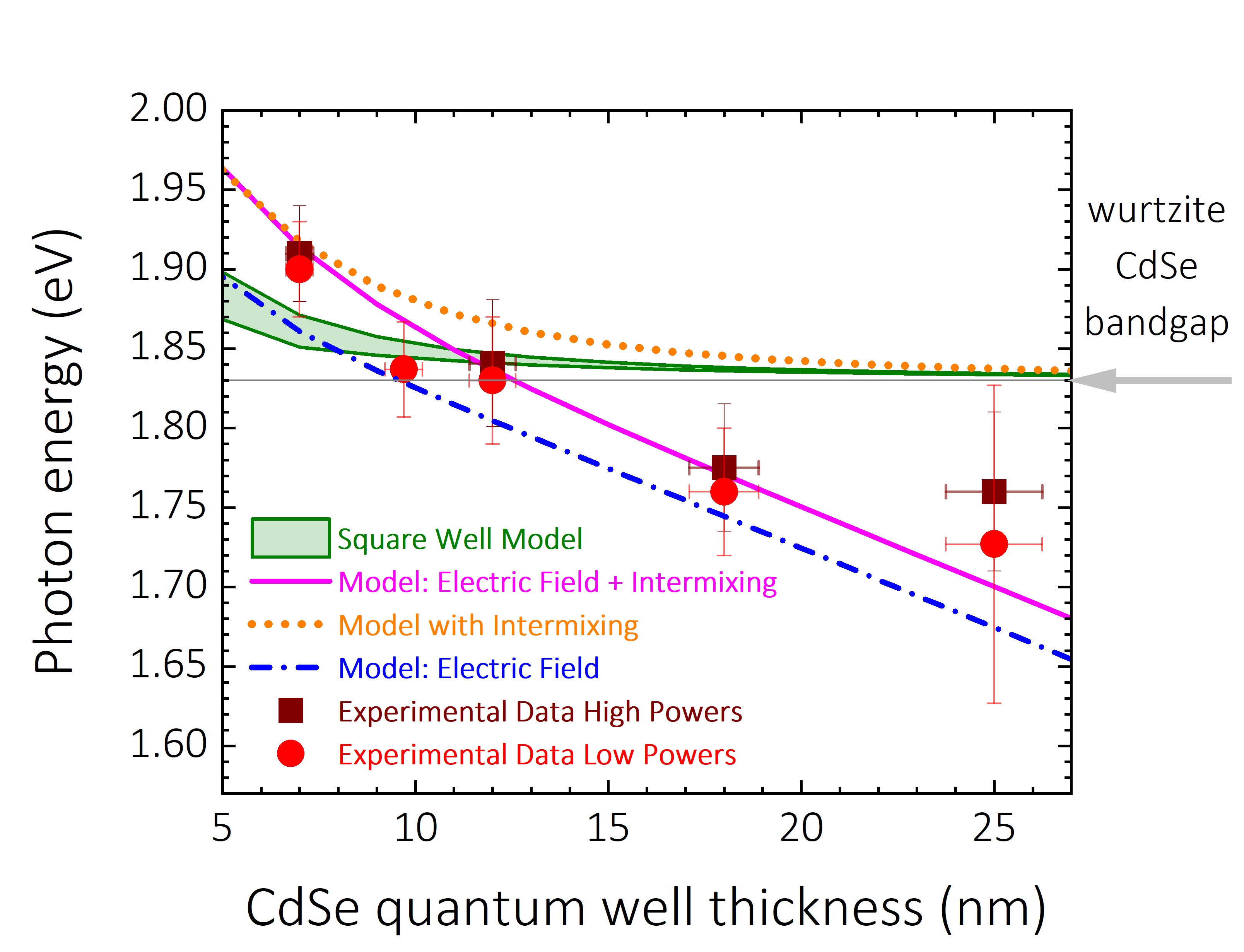}
\caption{\label{fig:energie}The comparison of the wurtzite CdSe QW photoluminescence energy with the results of simulation. The experimental data collected at $T=10\,$K for the excitation wavelength of $\lambda = 405\,$nm are represented by the red points (excitation power of $0.3\,$mW) and the brown squares (excitation power of $2.5\,$mW). The simulation results showing the expected recombination energies are displayed by the green area (simple model for the finite square quantum well), orange dotted line (model including intermixing effects), blue dotted-dashed line (model including built-in 14 MV/m electric field) and pink continuous line (model including both intermixing and 14 MV/m built-in electric field). The green area corresponds to MnSe valence band offsets ranging between 0.01 and 0.98, while all the other curves are calculated for MnSe valence band offset of 0.4 with respect to CdSe.}
\end{figure}

We will now focus on the analysis of the energy of the emission as a function of the QW width. We fit a Gaussian or an asymmetric Gaussian profile (for the 25~nm QW) to the peak corresponding to the QW signal to quantify its energy. Obtained energies are plotted as a function of the QW thickness as red points or brown squares in Fig.~\ref{fig:energie}. It can be easily noticed that for the thick QW the PL signal has lower energy than the band gap of wurtzite CdSe. On the other hand, for $7\,$nm QW the energy is significantly larger than it can be inferred from quantum confinement in a finite square quantum well, which will be described in the next paragraphs. Overall, the range of energies in which the QW signal is spanned is much larger than expected, and the dependence of the energy of the emitted photons on the QW thickness is much steeper than expected from a simple calculation of the energy states in a finite square quantum well. We posit that the two effects, namely the built-in electric field and intermixing of the layers, coexist and influence the energies of the PL from the QW. To validate the above (i), we present numerical modeling of the ground states of asymmetric QW with electric field and intermixing, (ii) study the excitation power dependence of the QW PL, and (iii) perform time-resolved PL experiments.

The calculation of the eigenstates and eigenenergies for 1D Schrödinger equation with an arbitrary potential is performed with Numerov algorithm \cite{Noumerov1924, Caruso2022}. In the context of the large uncertainty in the MnSe band gap, for the simulation, we assume it to be as high as $3.65\,$eV, based on reports concerning colloidal wurtzite nanoparticles \cite{Sines2010}. In such a way, it is (Cd,Mg)Se barrier height, which is known more precisely, that is more important in determining the ground state in QW. The details of the modeling can be found in the supplementary part \cite{sup}. The resulting outcome of the simulation is the energy of the photon emitted from the QW in the recombination process, which is roughly equal to the sum of the electron and heavy-hole confinement energy and the CdSe band gap. In this scenario, we neglect the exciton binding energy or any other many-body interactions. We take into account strong anisotropy of the effective mass of heavy-holes in wurtzite CdSe (and adopt heavy-hole mass along [0001] direction to be $1.17m_0$ and electron mass effective mass of $0.13m_0$ after \cite{Laheld1997}), treat heavy and light holes as non-degenerate \cite{Laheld1997} and neglect any potential coupling between heavy and light holes states induced by an electric field. We start our consideration with a simple square, finite quantum well. Band offsets of the MnSe-CdSe interface are not known; thus, we consider an array of all possible values of band offsets that still provide quantum confinement for both electrons and heavy-holes. The simulation result is represented as a green area in Fig.~\ref{fig:energie}, which shows all possible energies of photons emitted by the CdSe QW. It can be seen that the band offsets play a significant role in thin QWs only, whereas the uncertainty coming from this effect in thick QW is negligible. The dependence of the energy on the QW thickness is weak for thick QW, and the PL energy converges to the band gap of CdSe as expected. It is clear that a finite, square QW model is not enough to describe the observed experimental results represented by the red and brown dots.

\begin{figure}
\includegraphics[width=1\linewidth]{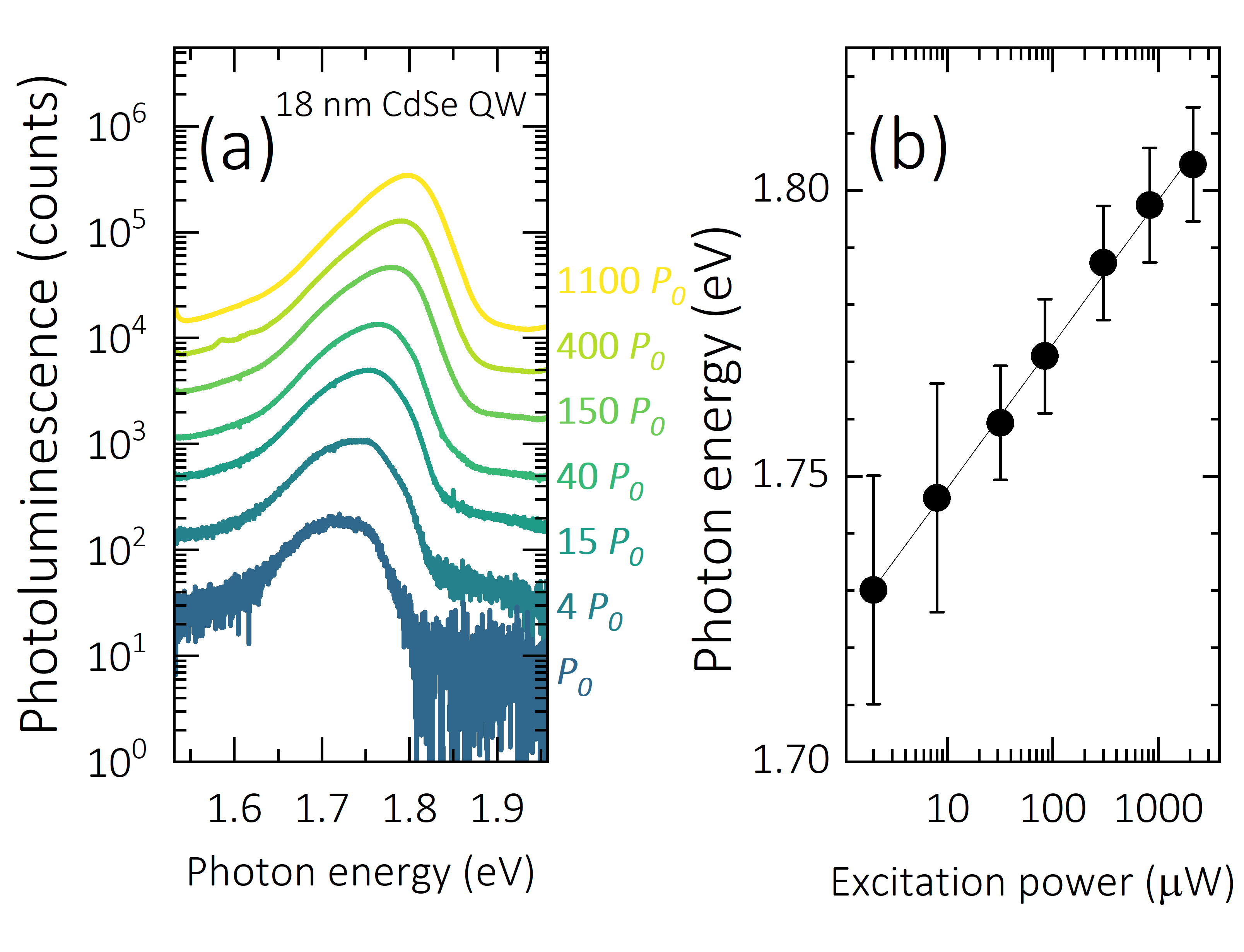}
\caption{\label{fig:power} (a) Dependence of $18\,$nm wide CdSe QW emission spectrum on the excitation power ($P_0=2\,\mu$W). The center of the peak associated with QW recombination shifts to high energies with increasing excitation power due to the increased screening of the electric field. (b) Peak position as a function of excitation power with a linear fit.}
\end{figure}

To account for the thin QW limit, we consider realistic imperfect interfaces affected by ions intermixing, the phenomenon, which is well-known from the earlier studies of II-VI QW and diluted magnetic semiconductors \cite{Gaj1994, Grieshaber1996, Kossacki1999, opion2022}. The intermixing makes nominally square potential rounded due to the migration of the ions, which effectively increases the band gap of the QW near the edges. Moreover, due to the epitaxial growth, this effect is different at the two interfaces of the QW.  To simulate it, an exponential profile of the QW \cite{Gaj1994} can be introduced. Such treatment elevates the energies of PL in the regime of thin QW as depicted by the orange dotted line in Fig.~\ref{fig:energie}, which is an example of a simulation for a chosen band offset. On the other hand, the effect of apparent sub-band-gap energy of PL from QW can be understood in terms of the built-in electric field that is relevant for wurtzite structures lacking inversion symmetry. Such electric field is significant for thick QW where the potential gradient builds up over a significant distance. We include the electric field in the QW potential landscape, which results in the slope of the bands in the real space coordinate. For simplicity, we assume that within the barriers the bands are flat but because a relatively small part of the wavefunction is localized outside QW it is an acceptable simplification. Such an electric field causes the energy to decrease with respect to the non-polar QW, and the effect is especially strong for thick QW as depicted by the dotted-dashed blue line in Fig.~\ref{fig:energie} \footnote{Excitation causes partial screening of the built-in electric field. Simulated curves in Fig.~\ref{fig:energie} correspond to an actual electric field of $10\,$MV/m, which was recalculated to the built-in electric field of $14\,$MV/m based on Fig.~\ref{fig:power} and Fig.~S2}. Finally, the interplay of the two: intermixing and electric field yields a good fit to the experimental data (continuous pink line in Fig.~\ref{fig:energie}).

Another observation confirming the presence of the built-in electric field in QW is the power dependence of QW emission. Higher power excitation triggers higher carrier concentration in the QW and stronger screening of the electric field. As shown above, screening of the electric field will tend to elevate the PL energy. Hence, for higher power excitations, we should observe PL peaks to shift towards higher energies. This can be seen as a difference between red and brown dots that have been collected for excitation power of $300\,\mu$W and $2.5\,$mW, respectively. An example of a systematic study of how energy changes with excitation power can be found in Fig.~\ref{fig:power}, where PL spectra of 18~nm QW for different excitation powers are presented. The peak of the QW signal changes energy by around 70~meV between the two extreme powers ($2 \mu$W and 2 mW). The shifts for QW thicknesses of 7, 12, and 25$\,$nm are $10\,$meV, $60\,$meV and $20\,$meV, respectively. Relatively small shift for the thickest QW may be related to the presence of the Coulomb interaction between the electrons and holes resulting in the exciton size smaller than QW width as well as possible contribution to the emission from excited states \cite{Muziol2022}. Extrapolation of photon energies of the QW PL singal to zero excitation power allows us to estimate that the built-in electric field in QW is $14\pm2\,$MV/m (see~\cite{sup}).

\begin{figure}
\includegraphics[width=1\linewidth]{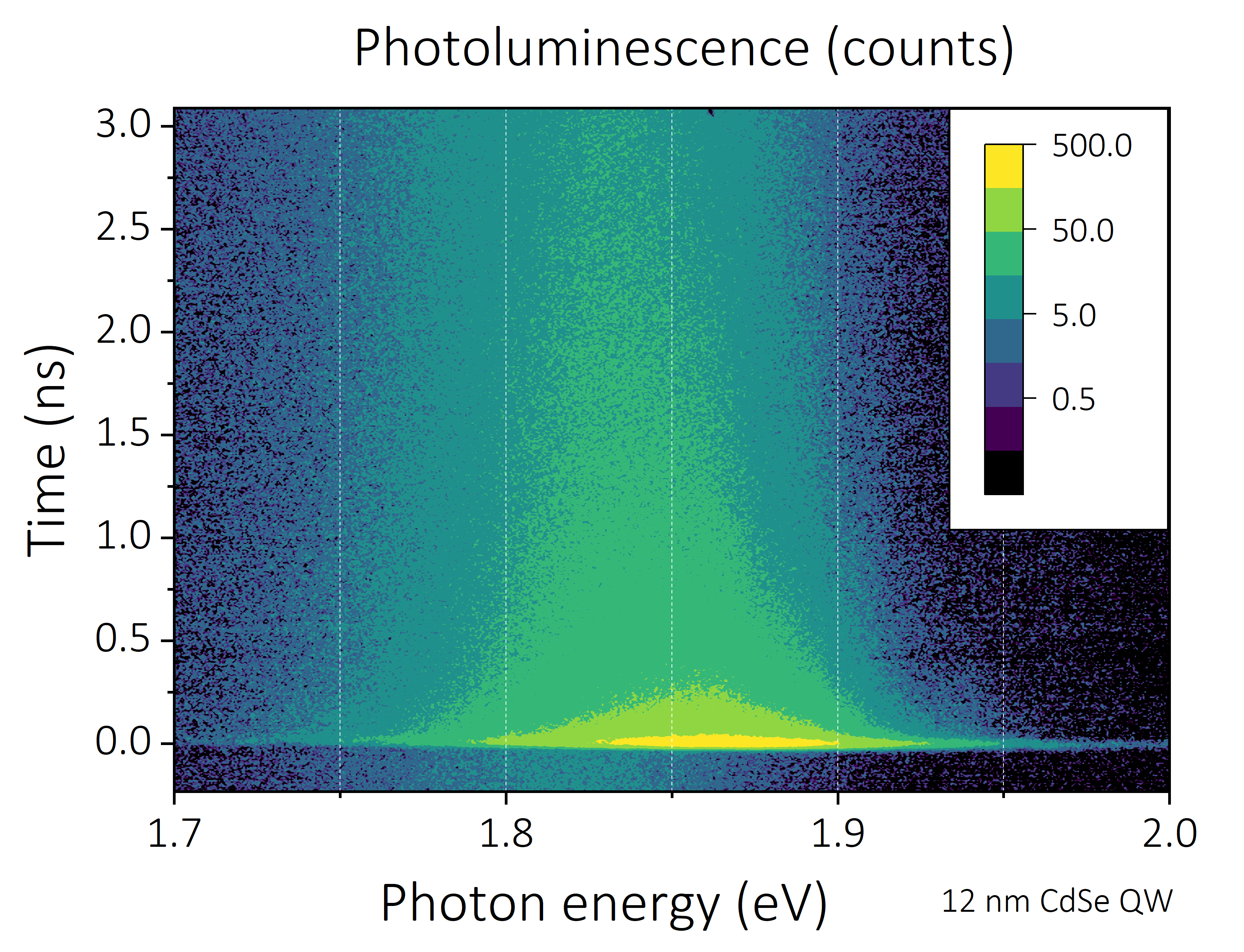}
\caption{\label{fig:timemap} Time dependence of PL intensity spectrum (colormap in the logarithmic scale) collected at $T=10\,$K for excitation with the pulsed laser of $\lambda_\text{exc}=432\,$nm and with the power of $P=0.1\,$mW. Both the decay of the PL signal with the decay time of the order of single ns and the shift to the lower energies with time are visible.}
\end{figure}

\begin{figure}
\includegraphics[width=1\linewidth]{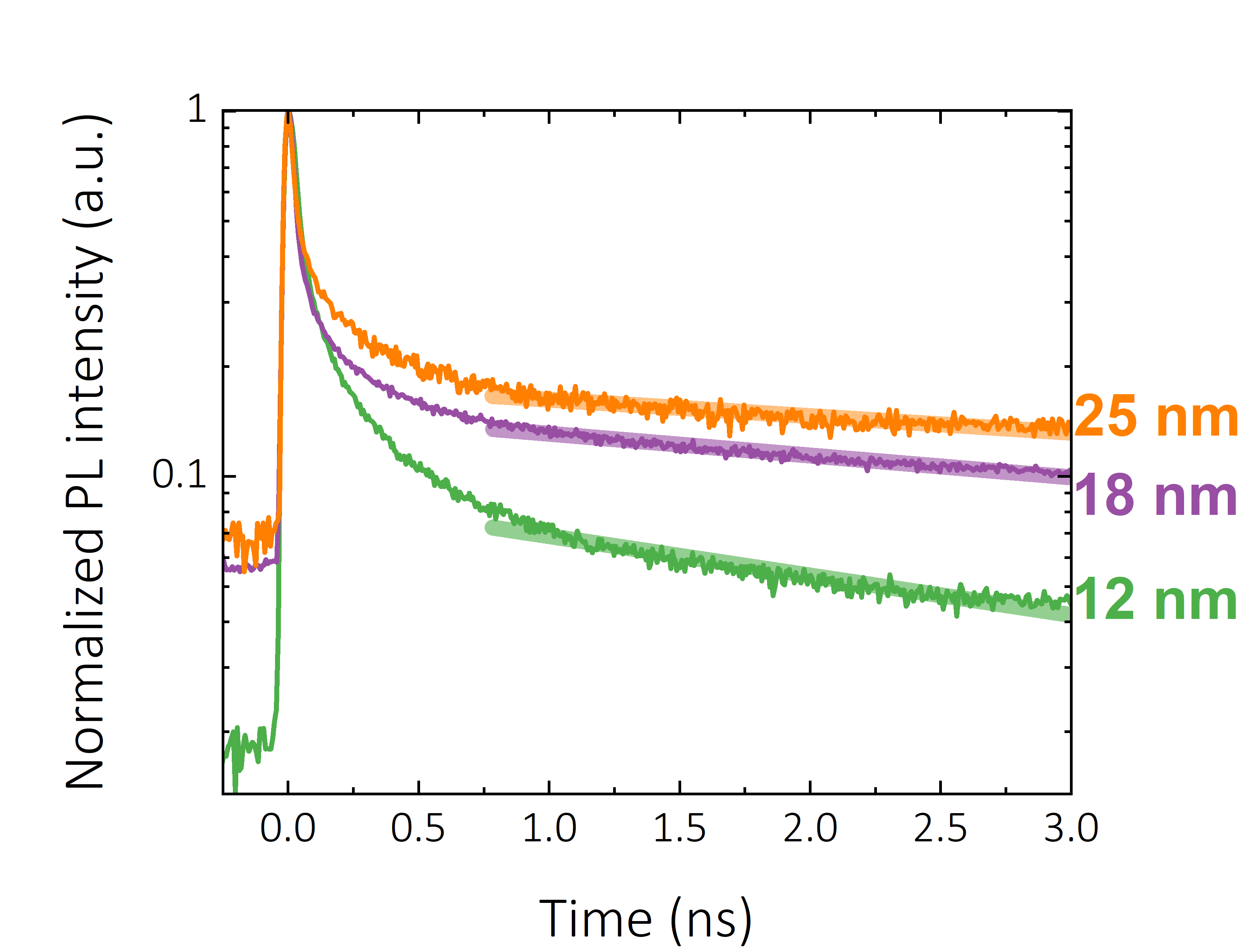}
\caption{\label{fig:dynamika}Dynamics of PL intensity averaged around the PL signal maximum for different QW thicknesses collected at $T=10\,$K for $\lambda_\text{exc}=432\,$nm excitation wavelength with the power of $P=0.1\,$mW. Thick straight lines represents linear fits performed for times larger than $0.75\,$ns.}
\end{figure}

\begin{figure}
\includegraphics[width=1\linewidth]{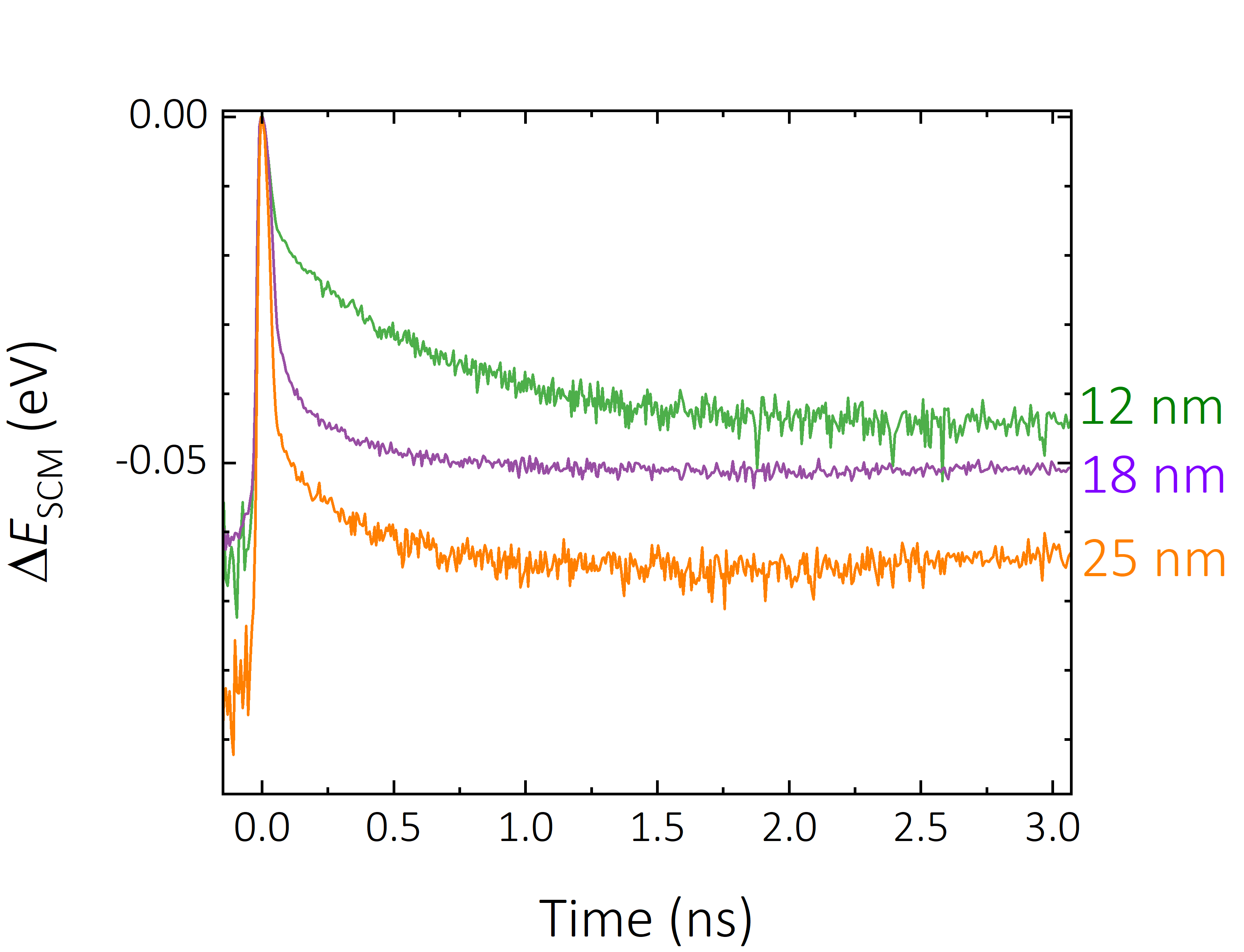}
\caption{\label{fig:scw}Relative changes of the center of mass of the spectra $\Delta E_\text{SCW}$ as a function of time fot excitation power of $P=100\,\mu$W, $\lambda_\text{exc}=432\,$nm and at $T=10\,$K.}
\end{figure}

\begin{figure}
\includegraphics[width=0.75\linewidth]{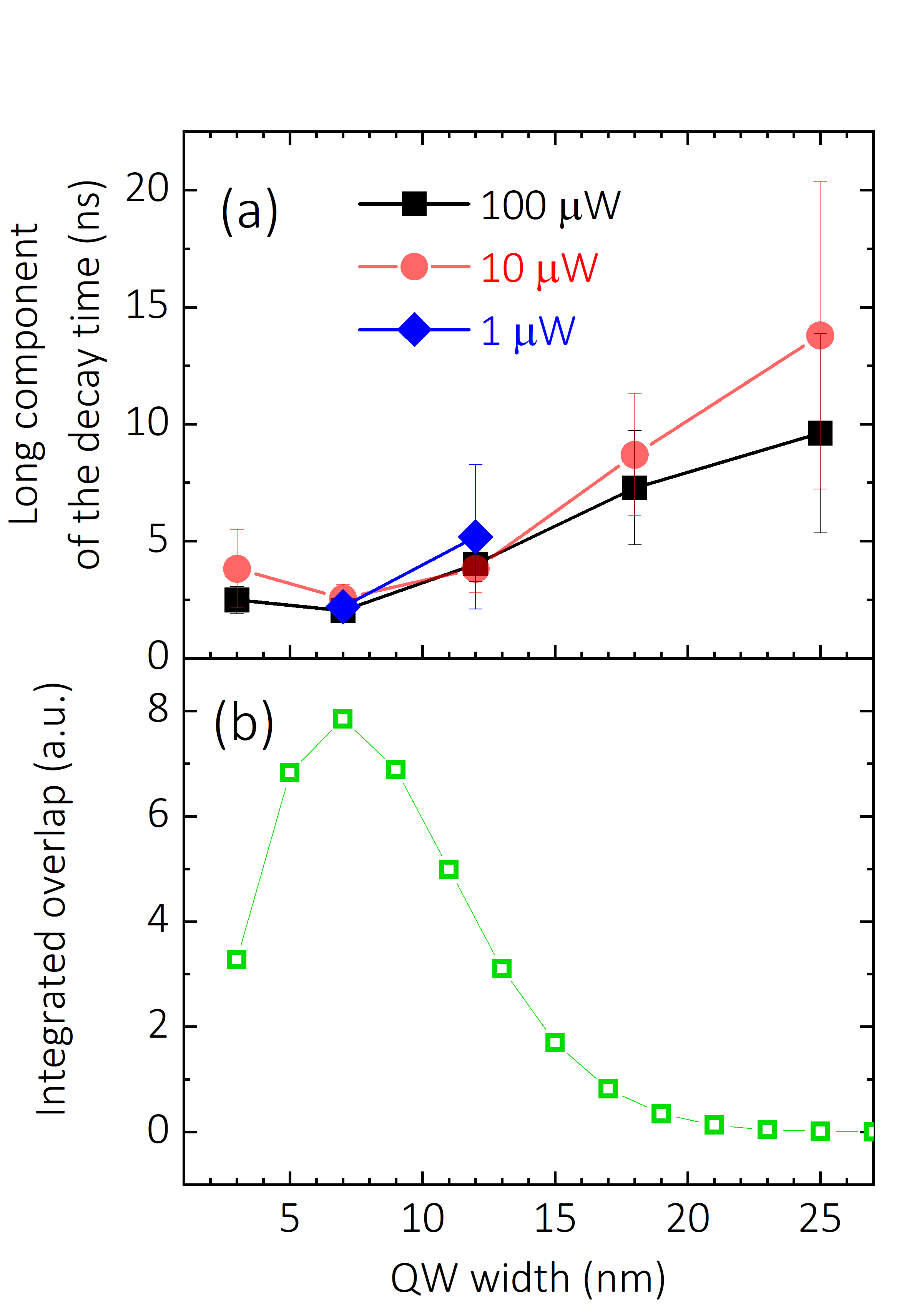}
\caption{\label{fig:tau}(a) Long component of the decay times determined from PL intensity dynamics (Fig.~\ref{fig:dynamika}) for different powers of excitation and different QW thicknesses. (b) Modeling of heavy-hole and electron wavefunctions overlap (for the parameters corresponding to the pink line in Fig.~\ref{fig:energie}).}
\end{figure}

Eventually, we present a time dependence of CdSe QW emission excited with a pulsed $432\,$nm laser and acquired using a Hamamatsu streak camera. The exemplary map of PL intensity as a function of energy and time is presented in Fig.~\ref{fig:timemap} for $12\,$nm QW. The spectral region is chosen so that the QW signal is approximately at half of the energy axis at the zero time. Both the spectral shift of the QW peak and the variation of the intensity following the excitation pulse are observed. The signal averaged in the spectral region around the peak maximum and plotted as a function of time for different QW thicknesses can be found in Fig.~\ref{fig:dynamika}. Additionally, for a more convenient comparison, the PL intensity is normalized with respect to the maximum intensity. The extreme cases agree with the expectations - for thick QW PL intensity remains large with respect to the initial values, whereas for the thinnest QW it drops much more rapidly, as it can be inferred from the separation of the electrons and holes due to the electric field. The dynamics of PL for all QW is clearly not mono-exponential (with a logarithmic Y axis, it would be visible as a linear slope). Because the concentration of the carriers in QW changes with time, so does the electric field due to the screening effect. A single decay time cannot be extracted from a fitting. However, we can perform such a linear fit after around $0.75\,$ns. Although in this range PL intensity changes less than an order of magnitude, the energy shift of the QW PL peak becomes insignificant (see also Fig.~\ref{fig:scw}), dynamics is much closer to mono-exponential than in the initial part so the long-range decay times can be estimated and treated as a valuable, systematically determined parameter differentiating QWs. Remarkably, minimal decay time corresponds roughly to the maximum overlap of the electron and heavy-hole states according to the modeling (Fig.~\ref{fig:tau}) with electric field and intermixing included. The decreased overlap and increased decay time for very thin QW originate from the fact that a significant part of the wavefunction is localized outside the QW.

Complementary observations are brought by the center of mass of the spectra $\Delta E_\text{SCW}$ analysed as a function of time. Relative changes of $\Delta E_\text{SCW}$ are plotted for three different QW thicknesses in Fig.~\ref{fig:scw}. At the time of the laser pulse excitation $\Delta E_\text{SCW}=0$ and due to the high carrier concentration the electric field screening is strong. Shortly afterwards $\Delta E_\text{SCW}$ gains larger magnitude for thicker QWs which is in agreement with the presence of the built-in electric field. 


In conclusion, we show a successful incorporation of epitaxial wurtzite MnSe layers into a low-dimensional structure, where it acts as an efficient potential barrier for carriers confined in a CdSe QW. We investigated QW emission in photoluminescence experiments for different QW thicknesses. We observed strong emission in the visible range. The results of such experiments were compared with the calculated emission in QWs with intermixing and electric field. Hence, we find a strong built-in electric field in such CdSe QW on the order of $14\,$ MV/m. This conclusion is corroborated with the experimental results of the excitation power dependence of QW emission and PL with temporal resolution. The extreme of a long component of the PL decay time and the calculated integrated overlap of holes and electrons coincide for the 7~nm QW. We also notice the influence of interface intermixing significant for the thinnest QWs. The common wurtzite crystal structure for CdSe and MnSe in the studied system suggests that a strong electric field is also present in MnSe. This makes wurtzite MnSe an attractive material, not only because of its potential altermagnetism, but also due to the prospective strong built-in electric field and its suitability as a building block for quantum wells.

\begin{acknowledgements}

We would like to thank Piotr Kosscaki, Tomasz Kazimierczuk, Kacper Oreszczuk, Mateusz Goryca, Carmine Autieri, Andrzej Golnik, Karel V\'{y}born\'{y} for helpful discussions and insightful questions, as well as Jan Pawłowski for AFM measurements. This research was funded in part by National Science Centre, Poland 2021/40/C/ST3/00168 and 2023/51/D/ST3/01978. J. S. acknowledges the support within the IDUB project No. BOB-661-1055/2025 financed by the University of Warsaw.

\end{acknowledgements}

\nocite{*}
\bibliography{bib}

\end{document}